\documentclass{aa}
\usepackage{graphicx}
\usepackage{epsfig,times}
\usepackage{color}

\begin{document}
 
\title{Nucleosynthesis in neutron-rich ejecta from Quark-Novae}
 
\author{Prashanth Jaikumar~\inst{1}
\and Bradley S. Meyer~\inst{2}
\and Kaori Otsuki~\inst{3}
\and Rachid Ouyed~\inst{4}}

\institute{
Department of Physics and Astronomy, Ohio University, Athens, OH 45701 USA\thanks{email:jaikumar@phy.ohiou.edu}
\and
Department of Physics and Astronomy,
Clemson University, Clemson, SC 29634 USA
\and 
Department of Astronomy and Astrophysics, Enrico Fermi Institute, University of Chicago, Chicago, IL 60637 USA
\and
Department of Physics and Astronomy, University of Calgary, 2500 University Drive NW, Calgary, Alberta, T2N 1N4 Canada
}

\date{Received <date>; accepted <date> }

\authorrunning{Jaikumar et al.}

\titlerunning{Nucleosynthesis in neutron-rich ejecta from Quark-Novae} 

\abstract{
We explore heavy-element nucleosynthesis by rapid neutron capture
(r-process) in the decompressing ejecta from the surface of a neutron
star. The decompression is triggered by a violent phase transition to 
strange quark matter (quark-nova scenario). The presence of neutron-rich large Z nuclei $(40,95)<(Z,A)<(70,177)$, the large neutron-to-seed ratio, and the low electron fraction $Y_e\sim 0.03$ in the decompressing ejecta present favorable conditions for the r-process. We perform network calculations that are adapted to the quark-nova conditions, and which mimic usual ($n-\gamma$) equilibrium r-process calculations during the initially cold decompression phase. They match to dynamical r-process calculations at densities below neutron drip ($4\times 10^{11}$ g cm$^{-3}$). We present results for the final element abundance distribution with and without heating from nuclear reactions, and compare to the solar abundance pattern of r-process elements. We highlight the distinguishing features of quark-novae by contrasting it with conventional nucleosynthetic sites such as type II
supernovae and neutron star mergers, especially in the context of
heavy-element compositions of extremely metal-deficient stars.
 }

\maketitle

\section{Introduction: {\sl astrophysical sites for the r-process}}

The r-process is of fundamental importance in explaining the origin of
many heavy neutron-rich nuclei ($A>90-100$)
(\cite{Burbidge}; \cite{Cameron}). While the astrophysical
conditions that lead to a successful r-process can be estimated, the
astrophysical site that can provide these conditions evades positive
identification. Recent comprehensive spectroscopic studies of
extremely metal-poor stars in the Galactic halo (\cite{Hill};
\cite{Cowan}; \cite{Sneden03}), as well as chemical evolution
studies (\cite{IshWan}; \cite{Ishi}; \cite{Argast04}), confirm that a) 
the r-process is a robust primary nucleosynthetic mechanism and b) 
core-collapse (type II) supernovae
(SNe) are favored over neutron star mergers (NSMs) as its astrophysical
site. However, a robust r-process model without
parameterization in the neutrino-driven wind of a core-collapse
supernova is, thus far, lacking (\cite{Qian03b}). Other astrophysical sites
explored in the context of a successful r-process include: prompt supernova 
explosions from small iron cores (\cite{Sumiyoshi}) or from O-Ne-Mg cores 
(\cite{Wanajo}), and neutron star mergers (\cite{FRT}). Although these are 
promising candidates, they face major theoretical and observational challenges,
as outlined below. In this paper, we explore decompressing neutron matter from 
the surface of a neutron star as an alternative or additional site for r-process
nucleosynthesis. Previously, decompression was attributed to tidal disruptions
in neutron star-neutron star or neutron star-black hole collisions
(\cite{LS}, \cite{SS}, \cite{FRT}), explosion of a neutron star below its
minimum mass (\cite{Sum}), or equatorial mass shedding
during spin-down of rapidly rotating supramassive neutron stars. 

\vskip 0.2cm

\noindent In this work, we consider a new decompression scenario: a violent nuclear-quark phase transition in the core of a neutron star that ejects neutron-rich matter at 
the surface: the quark-nova (\cite{KOJ}). Our main aim is to evaluate the 
quark-nova (QN) as a nucleosynthetic site and contrast it with more conventional alternatives, namely Type II SNe and NSMs, in the context of observations. We 
begin by highlighting important observations that can judge relative 
merits of the astrophysical sites studied so far.

\subsection*{r-process-enhanced metal-poor stars}

 The extremely metal-poor ([Fe/H] = -3.1) halo giant CS 22892-052 shows an 
overabundance of r-process nuclei compared to Fe-peak nuclei; e.g., Eu/Fe 
exceeds the corresponding solar values by factors of 30 or more (\cite{Sneden96}, \cite{Sneden03}). However, the relative element abundances in the range 
$56 < Z < 82$ (i.e., Ba through Pb) in CS 22892-052 follow the solar r-process 
pattern. This suggests that the (strong) r-process operates largely unchanged 
over the long history of Galactic chemical evolution, and is associated with 
a unique astrophysical process. Considerable effort
has therefore been invested in detailed modelling of core collapse SNe
and NSMs, accompanied by chemical evolution studies, in the hope of explaining the abundance observations and scatter plots of r-process-only elements such 
as Europium (Eu) (\cite{Argast00}, \cite{Argast04}). 

\vskip 0.2cm

\noindent NSMs occur at a much lower rate~\footnote{The coalescence rate for binary 
neutron stars is assumed to range from $3.10^{-4}$/yr to $10^{-6}$/yr 
(\cite{Belczynski}) with coalescence times that range from 100-1000 Myr
(\cite{Port}, \cite{Fryer99}).} than any type of SNe and so face problems in 
explaining the early onset (at times corresponding to [Fe/H]$\approx -2.5$) 
of correlations between the ratio of r-process elements [$r$/Fe] and Fe 
enrichment (\cite{Qian00}). Furthermore, neutron star mergers are unable 
to satisfactorily explain the associated scatter plots derived from observations 
of extremely metal-poor halo stars ([Fe/H]$<-3.0$) and thick disk stars ($-1.0<$[Fe/H]$<0.2$), which show that the scatter in [Eu/Fe] falls from $\sim$
2.0 dex at early times to $\sim$ 0.2 dex at later times (\cite{Argast04}).

\vskip 0.2cm

\noindent Comparative observations of r-element abundance patterns in metal-poor
stars such as CS 22892-052 and CS 31082-001 ([Fe/H] = -2.9) suggest that elements 
from Ba and above ($A>130$) follow the solar r-element pattern while those below 
$A\sim 130$ fall below the extension of this pattern. A similar conclusion is drawn
from the behaviour of the Strontium to Barium ([Sr/Ba]) ratio in a
comparison of r-process rich and r-process poor stars (\cite{Truran}). An 
explanation of this fact based on core-collapse SNe may originate in differing 
progenitor masses (\cite{Qian04})\footnote{Alternate sites for heavy-element 
r-processing include accretion-induced collapse (AIC) of a white dwarf into a neutron star 
in a binary, outflows in gamma-ray bursts (\cite{Surman}) and collapse of hybrid stars
of white dwarf dimensions (\cite{MP}).}. Intriguingly, a successful match to the solar
r-process pattern (particularly around the 3rd r-process peak ($A\sim
195$)), produced by SNe events associated with the low-mass
progenitors, seems to require an unusually massive ($M\sim
2.0M_{\odot}$) neutron star mass (\cite{Otsuki}, \cite{Sumi}). Such heavy neutron 
stars with canonical radius ($R\sim 15$km) would have large central densities 
provided the high density equation of state is not too stiff. They may be
prone to a nuclear-quark deconfinement transition at a later
stage. 

\vskip 0.2cm

\noindent Data from metal-poor stars point toward the distinct possibility of multiple r-process sites with varying astrophysical environments. A conclusive association of r-processing with core-collapse
SNe events awaits further spectral identification of newly-synthesized
r-process nuclei (\cite{Maz}) and detection of
gamma-rays from decays of r-process nuclei (\cite{MH},
\cite{Qian98}). 

\subsection*{r-process: theoretical challenges}

Aside from the challenges posed by abundance observations, there are hurdles on
the theoretical side as well. Progress in the modelling of type II
SNe and $\gamma$-ray bursts has led to the viable scenario of
``neutrino-driven winds" from nascent neutron stars (\cite{Woosley}; 
\cite{Takahashi};\cite{Qian96};\cite{Wanajo01};
\cite{Otsuki};\cite{TBM};\cite{Cardall}), where presumably the r-process is 
successful if model parameters are chosen appropriately. Unfortunately, values of 
model parameters extracted from core-collapse supernova simulations, such as the 
entropy per baryon, electron fraction and expansion timescale do not reproduce 
the observed solar abundance of r-process elements. One possible implication of 
this is that the
r-process has an additional site with different neutron exposure. The
prompt-explosion model of a collapsing O-Ne-Mg core does reproduce the
solar-like r-abundance pattern (\cite{Wana};\cite{Wanajo}) but an artificial 
enhancement of the shock-heating energy is needed to obtain the requisite physical conditions for a successful r-process.

\vskip 0.2cm

\noindent In neutron star mergers, the amount of mass ejected depends strongly
on the poorly-known high-density equation of state, with stiffer
equations of state predicting larger mass loss (\cite{Rosswog}). General relativistic considerations for the system dynamics, the sensitivity of the network
calculations to the electron fraction (which actually has a spread in
neutron star crusts), and the effects of neutrino transport on the
former during the coalescence phase present sources of uncertainty.
Furthermore, a universal description of all relevant nuclear data that would ensure good systematics remains out of reach, until reaction
cross-sections and decay half-lives of many unstable
exotic nuclei involved in the r-process network can be measured
accurately in rare-isotope experiments.

\vskip 0.2cm

 \noindent To overcome these challenges and explain the solar system
 composition in a satisfactory way requires new astrophysical input as
 well as better nuclear systematics. In this work, we wish to focus on
 the possibilities and implications of the decompression scenario with
 input parameters determined by the specifics of a quark-nova. This is
 not the first work to investigate r-process nucleosynthesis in
 decompressing neutron matter. Meyer (1989) and Goriely et al. (2004)
 have studied the r-process in the presence of exotic nuclei formed in
 matter resulting from the decompression of initially cold neutron
 star matter (\cite{latt77}), although no specific mechanism for
 decompression was proposed~\footnote{Alternate mechanisms for
 ejection of initially cold, decompressed neutron star matter were mentioned in
 the introduction.}. They found that improvements in radiative neutron
 capture rates by exotic nuclei close to the neutron drip
 line\footnote{In addition to the intermediate compound nucleus,
 direct electromagnetic transition to a bound final state, which is
 important for neutron-rich exotic nuclei, was included.} and fission
 probabilities of heavy neutron-rich nuclei lead to good agreement
 between the predicted and the solar abundance pattern for $A\ge 140$
 nuclei (see Figure 2 in \cite{Goriely}).  Those studies are the
 first consistent calculation of the nucleosynthetic composition of
 dynamically ejected material from the neutron star surface.

\vskip 0.2cm

 \noindent Interestingly, the initial conditions for decompression considered by
 Goriely et al. (2004) for a reasonably successful reproduction of the
 heavy-element r-process pattern are almost characteristic of the quark-nova
 explosion and its ejecta. After explaining the quark-nova scenario in
 section 2, we set up the decompression dynamics in the setting of a
 quark-nova in section 3. In section 4, we present results from our
 network calculations of the decompression scenario, and compare them
 with the solar r-process pattern. We consider cases of slow and fast 
decompression with and without nuclear heating from $\beta$-decays. 
In section 5, we compare and contrast the QN with NSM and type II SNe as astrophysical sites for the r-process, particularly in the context of Galactic chemical evolution. We address observational signals of the r-process event such as $\gamma$-rays from the radioactive decay of heavy r-process nuclei. Our conclusions and a discussion of certain aspects that have not been covered in this work, and which require further study, are presented in section 6.
 
\section{Quark-nova: {\sl nuclear-quark phase transition}}

It has long been thought that the center of neutron stars may be dense enough that 
nuclear boundaries dissolve and a phase transition to quark matter occurs (\cite{Itoh};\cite{Bodmer}). Up and down quarks would then preferentially convert to strange quarks (other heavier flavors can be neglected) in order to reduce Pauli 
repulsion by increasing flavor degeneracy. This idea was formally stated as a conjecture (\cite{Witten}) that strange quark matter is the true ground state of strongly 
interacting matter at zero pressure. Even if true only at finite pressure (\cite{Farhi}), its consequences for neutron stars are sufficiently intriguing that several possibilities for such a phase conversion have been examined (\cite{Alcock}). One attractive possibility is that nuclear matter undergoes a phase transition to two flavor (up and down) quark matter, which then undergoes weak equilibrating reactions to form three flavor quark matter (\cite{Olesen};\cite{Cheng}). The advantage is that this does not require the improbable simultaneous and spontaneous conversion (via weak interactions) of several neutrons within a small volume. In addition, the Coulomb-barrier-free absorption of neutrons can enlarge the quark phase. If Witten's conjecture holds, the whole neutron star is essentially converted to strange quark matter, thus forming a quark star.

\vskip 0.2cm

\noindent There are various ways in which this conversion can be triggered. A
rapidly rotating neutron star is slowed down principally by magnetic
braking (and additionally by energy loss through gravitational waves),
thereby reducing the centrifugal force and increasing the central
pressure. The probability of triggering a first order transition by
forming a tiny lump of two flavor quark matter (quantum nucleation) is
exponentially sensitive to the central pressure (\cite{Bombaci}). This makes the transition much more likely as the star spins down. Material accumulated from a fallback disk around a cooling neutron star can also force the conversion. Even the
Bondi accretion rate of $10^{21}$ g/sec from the interstellar medium
can lead to a mass increase of up to 0.1$M_{\odot}$ within a million
years which can be sufficient to compress the star beyond the minimum
density for the phase transition. Microscopic pathways for the phase conversion include the
clustering of $\Lambda$-baryons at high density and seeding by
energetic cosmic neutrinos or strangelets (\cite{Alcock}). Deconfinement of quarks can also occur as early as the protoneutron star stage after a supernova explosion if the central density is large enough. However, in this case, neutrino trapping in
the hot and dense interior can push the transition density higher,
delaying the collapse to the point where a black hole might be formed
instead (\cite{Prakash}). 

\vskip 0.2cm

\noindent In this work, we do not address in detail the mechanism by which the
conversion is triggered, but clearly it is important to determine the
likelihood of such a conversion, since it is linked to the frequency
of the event that generates the r-process elements in our model calculation. Yasutake et al. (2005) have determined that the evolutionary transition from rapidly
rotating neutron stars to strange stars due to spin down can lead to
an event rate of $10^{-4}-10^{-6}$ per year per Galaxy. A more
comprehensive analysis employing different equations of state (the
simplistic Bag model was employed in Yasutake et al. (2005)) and a
variety of initial conditions at neutron star birth finds a similar event rate (\cite{Staff}). In light of the limited work on the above topic, we will
assume that spin-down is responsible for the phase transition. 

\vskip 0.2cm

\noindent In the quark-nova picture, (\cite{ODD}; \cite{KO}; \cite{KOJ}) the 
core of the neutron star that undergoes the phase transition to the quark phase 
shrinks in a spherically symmetric fashion to a stable, more compact strange 
matter configuration faster than the overlaying material (the neutron-rich
hadronic envelope) can respond, leading to an effective core collapse.
The core of the neutron star is a few kilometers in radius to begin
with, but shrinks to 1-2 km in a collapse time of about 0.1 ms
(\cite{LBV}). The gravitational potential energy thus
released is converted partly into internal energy (latent heat of
phase transition) and partly into outward propagating shock waves
which impart kinetic energy to the material that eventually forms the
ejecta. The temperature of the quark core thus formed rises quickly to
about 10 MeV since the collapse is adiabatic rather than isothermal
(\cite{Gentile}). In \cite{KOJ}, the core
bounce was neglected, and neutrinos emitted from the conversion to
strange matter were assumed to transport the energy into the outer
regions of the star, leading to mass ejection. With
neutrino-driven mass ejection, most of the neutrinos that can
escape the core lose their energy to the star's outer layers of
neutron matter in the form of heat. Consequently, mass ejection is
limited to about $10^{-5}M_{\odot}$ for compact quark cores of size
(1-2) km. More importantly, the neutrinos raise the electron fraction 
in the ejected layers by converting neutrons to protons, with
a resultant $Y_e\sim 0.5$. This is unfavorable for producing the
distinct peaks of the r-process. In fact, simulations of core collapse
induced by a nuclear-quark phase transition (\cite{Fry})
have shown that for $Y_e\sim 0.5$, the bulk of the ejecta is helium,
although some iron and r-process elements are produced. 

\vskip 0.2cm

\noindent A more attractive possibility is that of core bounce, accompanied by
outward propagating shocks that can impart sufficient kinetic energy
to the outer layers, including the crust of the star. Since most of
the neutrinos will remain trapped within the advancing front of
strange matter as it engulfs the star (\cite{KOJ}), the
$Y_e$ of the ejected material is small, and allows for an effective
r-process. It is important to note that only a fraction of neutrinos
from weak processes in the {\it (u,d,s)} core can escape before the
entire star converts to strange matter. This assumes that the {\it
(u,d,s)} contamination front moves at supersonic speeds (detonation)
rather than by combustion (deflagaration). Studies by Horvath \& Benvenuto (1998) confirm that in a detonation regime, the star would
rapidly convert to quark matter within a time 0.1ms. More recent work,
assuming realistic quark matter equations of state, argues for strong
deflagration (\cite{Drago}), but this too may be preceded
by a compression shock (\cite{LBV}) in the nuclear region
that can expel surface material.  

\vskip 0.2cm

\noindent Although an accurate study of the
shock propagation and the advancing quark conversion front would require detailed
modelling of neutrino transport, as a first
approximation, we neglect neutrino transport at core bounce, since
only about 1\% ($10^{51}$ ergs out of $10^{53}$ ergs released in the
phase conversion) of the energy leaves the core in the form of
neutrinos over typical timescales of mass ejection 
(due to long diffusion times of neutrinos in the hot quark
core). The ejected mass fraction is expected to be larger (about
$10^{-2}M_{\odot}$) than the previous case which neglected core bounce
. This is analogous to the mechanism of mass ejection from phase
transitions in a supernova as considered by Fryer \& Woosley (1998),
which can cause considerably more mass ejection due to an outgoing
shock wave from the supersonic collapse of the strange matter
core. The main difference from their work is that in a quark-nova, the neutrinos remain trapped in the hot quark core as it grows and engulfs the star, and they cannot
increase the $Y_e$ of the ejected matter significantly (\cite{KOJ}).

\vskip 0.2cm

\section{Decompression of the ejecta}

In a quark-nova, the decompression is powered by shock waves, and it
is possible that shock heating of the ejecta up to temperatures of
$10^{10}-10^{11}$K increases the $Y_e$ due to positron capture by the
neutron-rich material if it is not expanding faster than weak
interaction timescales. Thus, a rapid ejection of the material is
essential for efficient r-process nucleosynthesis. Apart from being
neutron-rich, the QN ejecta is also rich in exotic nuclei with $56 < A
< 118$ which come from the outermost layers of densities at and below
neutron drip (\cite{BPS}). Thus, electrical neutrality demands that there should be a spread in $Y_e\approx$ 0.03-0.2 in the crust. Since our calculations are
performed for a particular initial $Ye$, our results are to be interpreted as 
corresponding to a mass-averaged $\langle Y_e\rangle$, which depends on the ejected mass. We comment on this sensitivity of the yield to ejected mass (through $\langle Y_e\rangle$)
at the end of section 4.

\vskip 0.2cm

\noindent It is also possible that the
initial temperature of the ejecta ($T_9^i$) is variable, depending on
whether the transition to quark matter happens within seconds of the
protoneutron star formation (in which case $T_9^i\sim$ 10) or much
later ($T_9^i \sim$ 0.1 or smaller). The escaping neutrinos will also
increase the temperature of the neutron-rich material. As shown by
Meyer (1989) and Rosswog et al. (2000), the conditions for
nucleosynthesis in the decompression scenario are largely independent
of the choice of initial temperature within a range $T_9\approx 0-1$,
since in all cases, the expanding material heats up to r-process like
temperatures by $\beta$-decays, so we choose $T_9^i\approx 1$ or
smaller as our initial temperature. Since mass ejection is largely
unaffected by internal heating from neutrino absorption
(\cite{Rossw}), the temperature evolution of the ejecta
is subsequently determined by adiabatic expansion and nuclear
reactions (specifically $\beta$-decays when they become allowed). The
dynamical ($n-\gamma$) equilibrium calculations employed here allow us
to study this temperature evolution.

\subsection*{Initial conditions for decompression}

For the matter that is eventually ejected from the neutron star, the
compositions above and below neutron drip, as well as the nuclear
models used, are taken from the work of Meyer \& Schramm (1988), and
Meyer (1989). This improves upon the results of the compressional liquid
drop model pioneered by Baym, Bethe, \& Pethick (1971), and Lattimer
 et al. (1985) by including nuclear deformations and shell
effects. For the equation of state for dense, $\beta$-
equilibrated matter described in Meyer (1989) and Goriely et al.
(2004), the composition at $\rho\simeq 10^{14}$ g/cc is characterized by
an initial electron fraction $Y_e$=0.03, corresponding to a
Wigner-Seitz cell made by a $Z$=36, $N$=157 nucleus. However, our network
calculations are performed with $Y_e$ as a parameter for reasons explained
previously. The details of the computational methods are given in the 
following section. 

\vskip 0.2cm

\noindent The expansion of this dense ejecta (including $\beta$-decays) is
followed down to neutron drip density using the network calculation of
Meyer (1989), where the matter ends up distributed over a wide range
of elements from $Z\sim 40-70$. At this stage, the full r-process
reaction network is coupled to obtain the final abundance
distribution, as well as the evolution of temperature, entropy and
electron fraction. An important parameter other than $Y_e$ is the expansion
timescale for the ejecta, which is the time taken to evolve from a
density of $\rho\simeq 10^{14}$g/cc to $\rho\simeq
10^{11}$g/cc. Assuming that the ejecta is on an escape trajectory and
expands as a multiple of the free-fall timescale in a spherically
symmetric, non-interacting manner, we obtain a timescale on the order
of 0.01-0.1 milliseconds. 
\vskip 0.2cm

\noindent Since neutrinos are trapped within the quark
core on account of their long diffusive timescales of 0.1s, the
expansion is fast enough that $Y_e$ is not much changed by neutrinos,
as explained previously. Nevertheless, the results are expected to be
sensitive to the expansion timescale, so we vary this parameter from
0.01-1 milliseconds to test the sensitivity of our abundance
distribution to this parameter. In their calculations, Goriely et
al. (2004) find that the final abundance pattern can be significantly
different from solar for expansion timescales $\tau<3$ms, since not
all neutrons are captured. In the quark-nova scenario, the rapid
conversion to strange quark matter constrains us to explore a bounded
range of timescales $0.01<\tau ({\rm ms})<0.1$. For the sake of
comparison with results in Goriely et al . (2004), we also perform one set 
of calculations with $\tau\sim 1$ms, which would correspond to a slow expansion, 
but is not likely to be the case in the quark-nova scenario. The following 
section contains the results for heavy-element nucleosynthesis in the 
decompression scenario. A comparison to the solar abundance pattern in each case 
is also presented.

\section{r-nuclei abundance distribution}

We use the Clemson University nucleosynthesis code (\cite{Jordan})
to explore the nuclear yields that result from the decompression of matter ejected from a quark-nova.  This particular network includes
species from neutrons and protons up to uranium and all isotopes between
the proton-drip line to several nuclides beyond the neutron-drip line
for each element.  We choose to extend the network beyond the neutron-drip
line because some of our expansions reach such low temperatures that
neutron captures are able to branch across neutron-unstable
species in some cases. The last species in the network $Z$=$92$, $A$=$276$
is assumed to fission into two fragments--one with $Z$=$40$ and one with
$Z$=$52$. This provides strong fission cycling of the r-process flow in certain
cases. 

\vskip 0.2cm

\noindent We follow the matter beginning at $Y_e$=$0.03, T_9$=$1$ and an entropy per nucleon $s/k_B$=$1$.  We compute the temperature and density history of the material 
by imagining a spherical chunk of matter of radius $R$ decompressing
due to its own pressure $P$. In the absence of gravity (free fall), the momentum equation,
under the assumption of spherical symmetry, reads
\begin{equation}
\frac{dv}{dt} = -\frac{1}{\rho}\frac{\partial P}{\partial r}\quad,
\label{eq:momentum_eq}
\end{equation}
where $v$ is the velocity of a radial shell,
$\rho$ is the matter density, and $r$ is the radial coordinate of one
of the shells.  For simplicity, we assume a uniform density in any chunk
of expanding matter (at a particular instant) so that
\begin{equation}
\frac{\rho}{\rho_0} = \left( \frac{R_0}{R} \right )^3\quad,
\label{eq:density}
\end{equation}
where $\rho_0$ and $R_0$ are the density and and radius of the chunk
at the beginning of the calculation. We approximate
Eq. \ref{eq:momentum_eq} as
\begin{equation}
\frac{d^2R}{dt^2} = \frac{\alpha}{\rho}\frac{P}{R} = \alpha\frac{c_s^2}{R}\quad,
\label{eq:simple_momentum_eq}
\end{equation}
where $\alpha$ is a constant that accounts for deviations from
our simple model while also determining the expansion timescale, 
and $c_s$ is the sound speed. The pressure gradient causes the chunk of matter to accelerate outwards. This increases the radius, decreases the density and
pressure and hence, the acceleration. This means that after an initial
acceleration, the material tends to reach a coasting speed, which is
typically a few percent the speed of light in our calculations.

\vskip 0.2cm

\noindent We compute the pressure from routines available in
the nucleosynthesis code, as described in Meyer \& Brown (1997). For each timestep in the calculation, we use Eq. \ref{eq:simple_momentum_eq}
to find the radius of the chunk based on information from the previous timestep.
We compute the density from Eq. \ref{eq:density}.  We then guess the
temperature, solve the nuclear network for the changes in the abundances
over the timestep, compute the change in the entropy due to nuclear reactions,
and then invert the entropy to find the temperature.  If the resulting
temperature differs from our guess, we modify our guess and repeat the
procedure until we find agreement between our guess and the resulting
temperature.  We then proceed to the next step. 

\vskip 0.2cm

\noindent Nuclear reactions out of equilibrium generate entropy and consequently, lead to heating of the material during the expansion.  We assume any energy released by reactions is fully deposited locally.  There is a difficulty with this assumption. The primary heating in our calculations comes from beta decays. A significant fraction (typically $\sim 60\%$ for the
neutron-rich heavy nuclei in the r-process) of the energy from a beta decay
is released as a neutrino, which would escape the chunk without depositing
its energy.  We do not yet have the details of the relative energy carried
away by neutrinos for all the nuclei in the network; thus, to explore this
issue, which turns out to have significant consequences for our results,
we study two extemes, namely, one in which we assume all energy generated
by nuclear reactions was deposited locally and one in which we neglect energy
released by the reactions.  The former case certainly overestimates the
entropy generation by nuclear reactions because it neglects neutrino losses.
In the latter case, the entropy during the expansion is constant.

\vskip 0.2cm

\begin{figure}[ht!]
\centerline{\includegraphics[width=0.5\textwidth,angle=0]{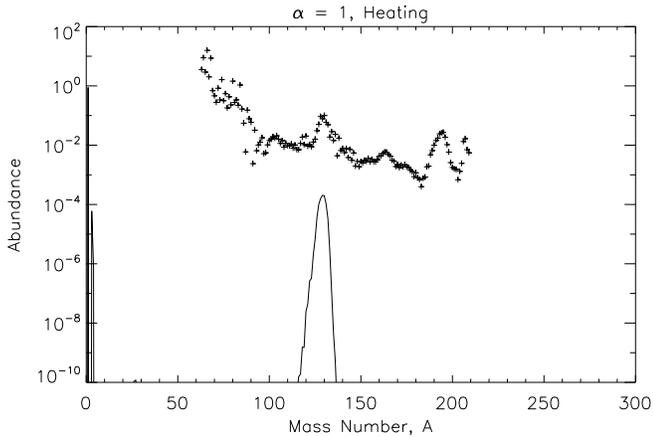}}
\caption{
 The final abundance distribution as a function of mass number for
 the \protect{$\alpha = 1$} calculation in the case in which heating from
 nuclear reactions was included. (See text below for details)}
\label{fig:ya_alpha=1_heat}
\end{figure}

\begin{figure}[ht!]
\centerline{\includegraphics[width=0.5\textwidth,angle=0]{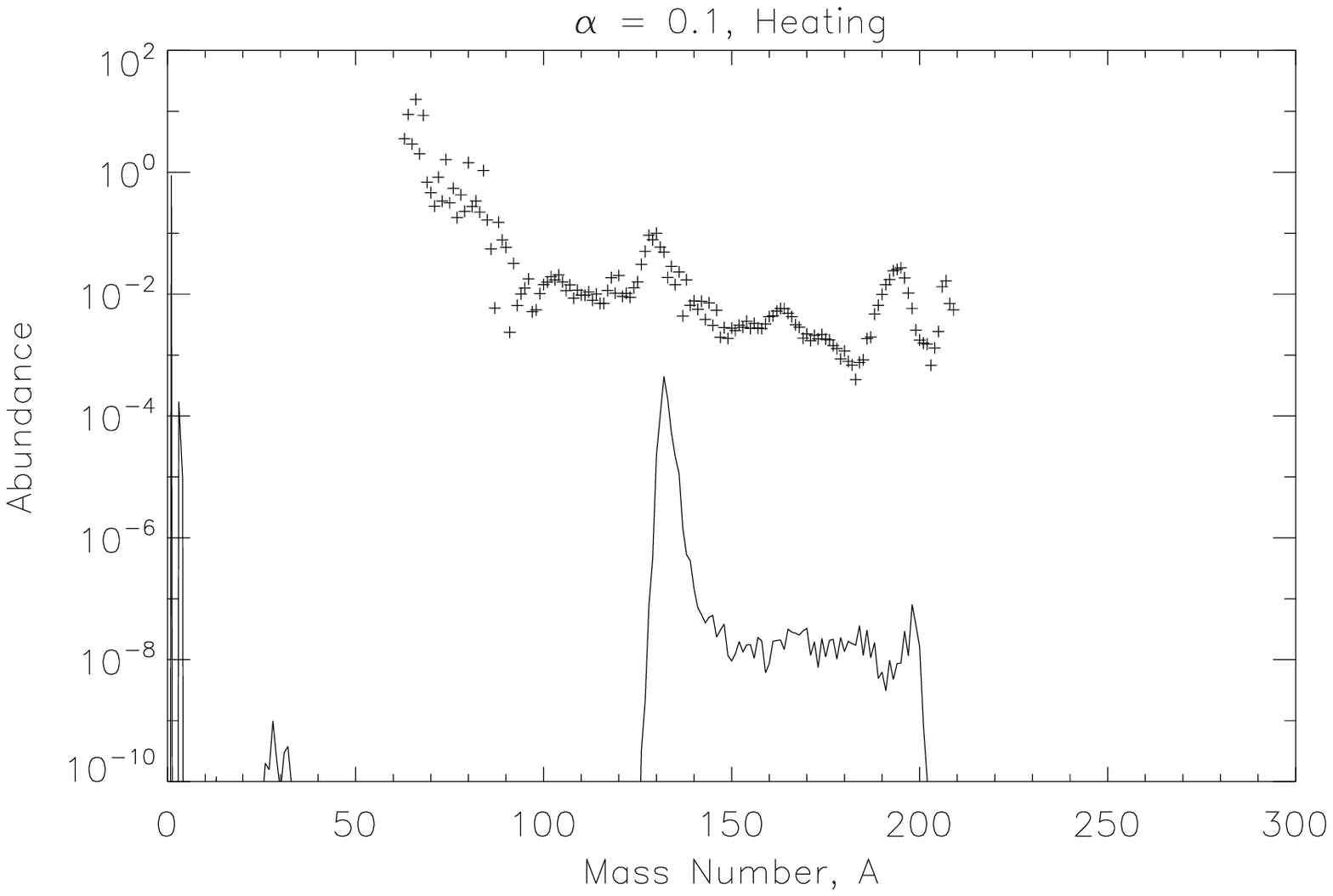}}
\caption{
 The final abundance distribution as a function of mass number for
 the \protect{$\alpha = 0.1$} calculation in the case in which heating from nuclear
 reactions was included. }
\label{fig:ya_alpha=0.1_heat}
\end{figure}

\begin{figure}[ht!]
\centerline{\includegraphics[width=0.5\textwidth,angle=0]{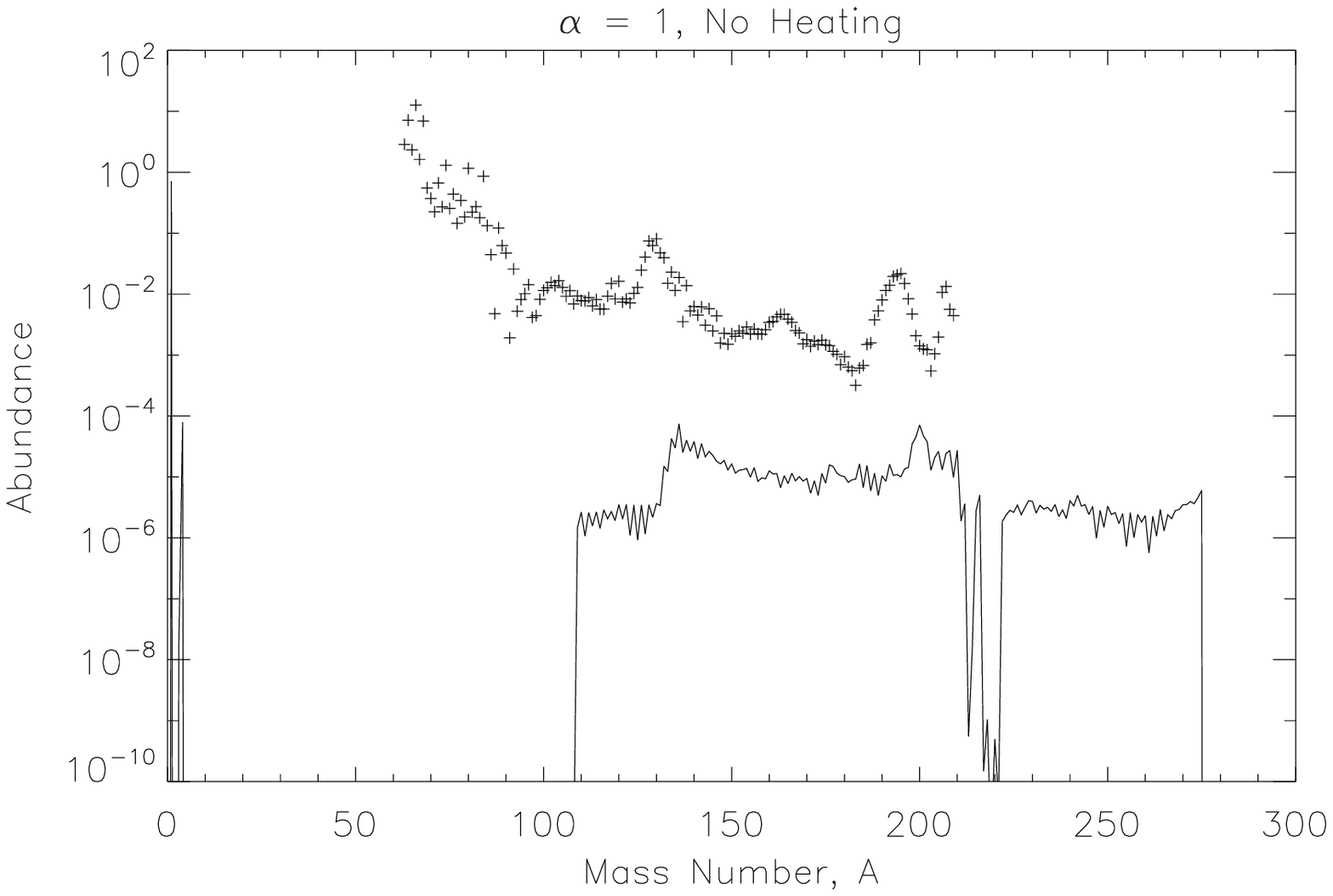}}
\caption{
 The final abundance distribution as a function of mass number for
 the \protect{$\alpha = 1$}
 calculation in the case in which heating from nuclear
 reactions was not included. 
}
\label{fig:ya_alpha=1_no_heat}
\end{figure}

\begin{figure}[ht!]
\centerline{\includegraphics[width=0.5\textwidth,angle=0]{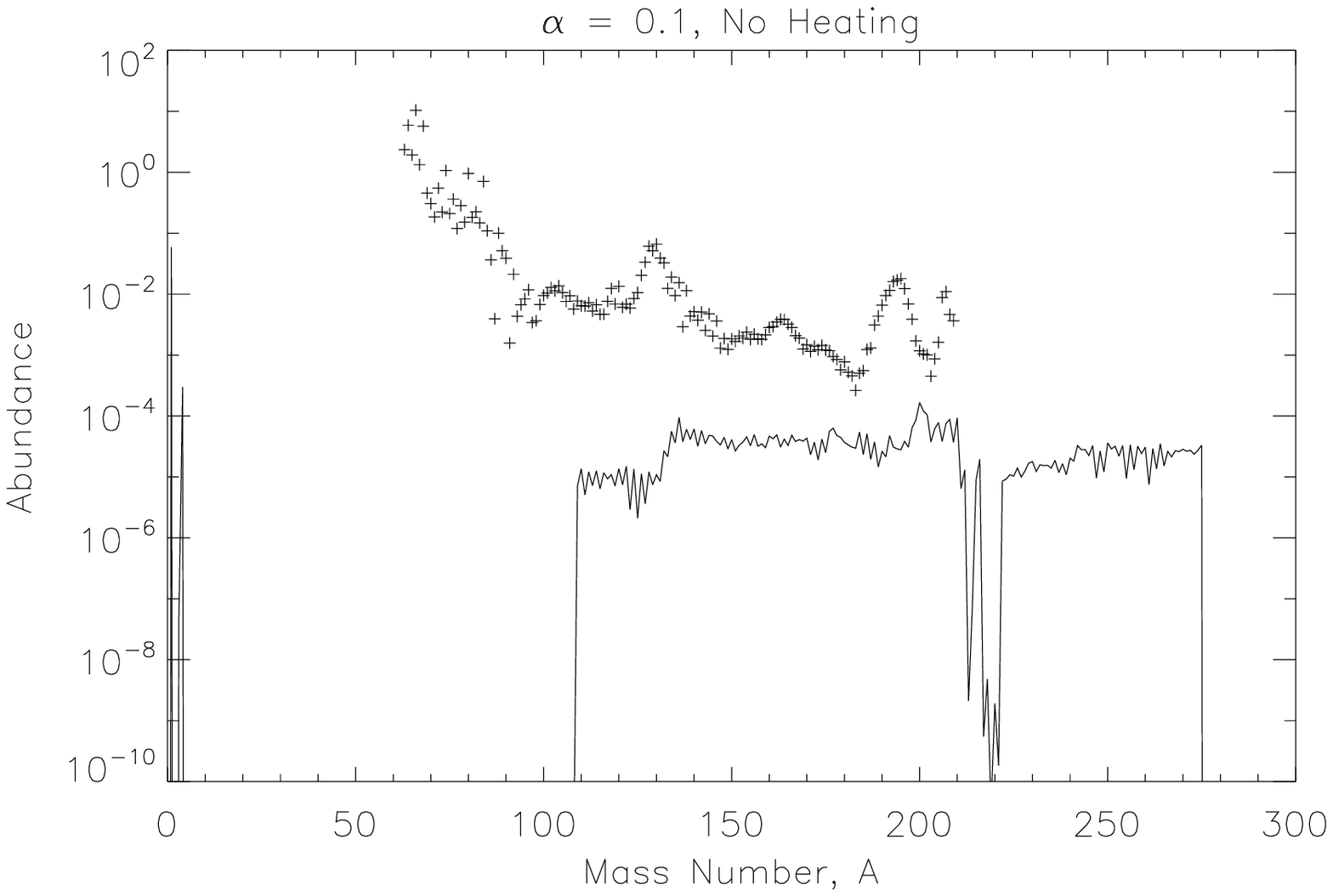}}
\caption{
 The final abundance distribution as a function of mass number for
 the \protect{$\alpha = 0.1$}
 calculation in the case in which heating from nuclear
 reactions was not included.}
\label{fig:ya_alpha=0.1_no_heat}
\end{figure}

\noindent We focus on four particular decompression calculations.  The calculations
have $\alpha$=$1$ (corresponding to an expansion timescale $\tau\sim 0.01$ms) or $\alpha$=$0.1$ ($\tau\sim 0.05$ms) and either include full or zero energy deposition from the nuclear reactions. $R_0$=$10$ meters and $\rho_0$=$10^{14}$g/cc in all cases. The final abundances resulting from these calculations are shown as a function of mass number $A$ in Figures \ref{fig:ya_alpha=1_heat}-\ref{fig:ya_alpha=0.1_no_heat}.
In each case, for comparison,the observed solar r-process abundances (Kappeler et al. 1989) are shown arbitrarily scaled as plus signs.

\begin{figure}[ht!]
\centerline{
  \includegraphics[width=0.5\textwidth,angle=0]{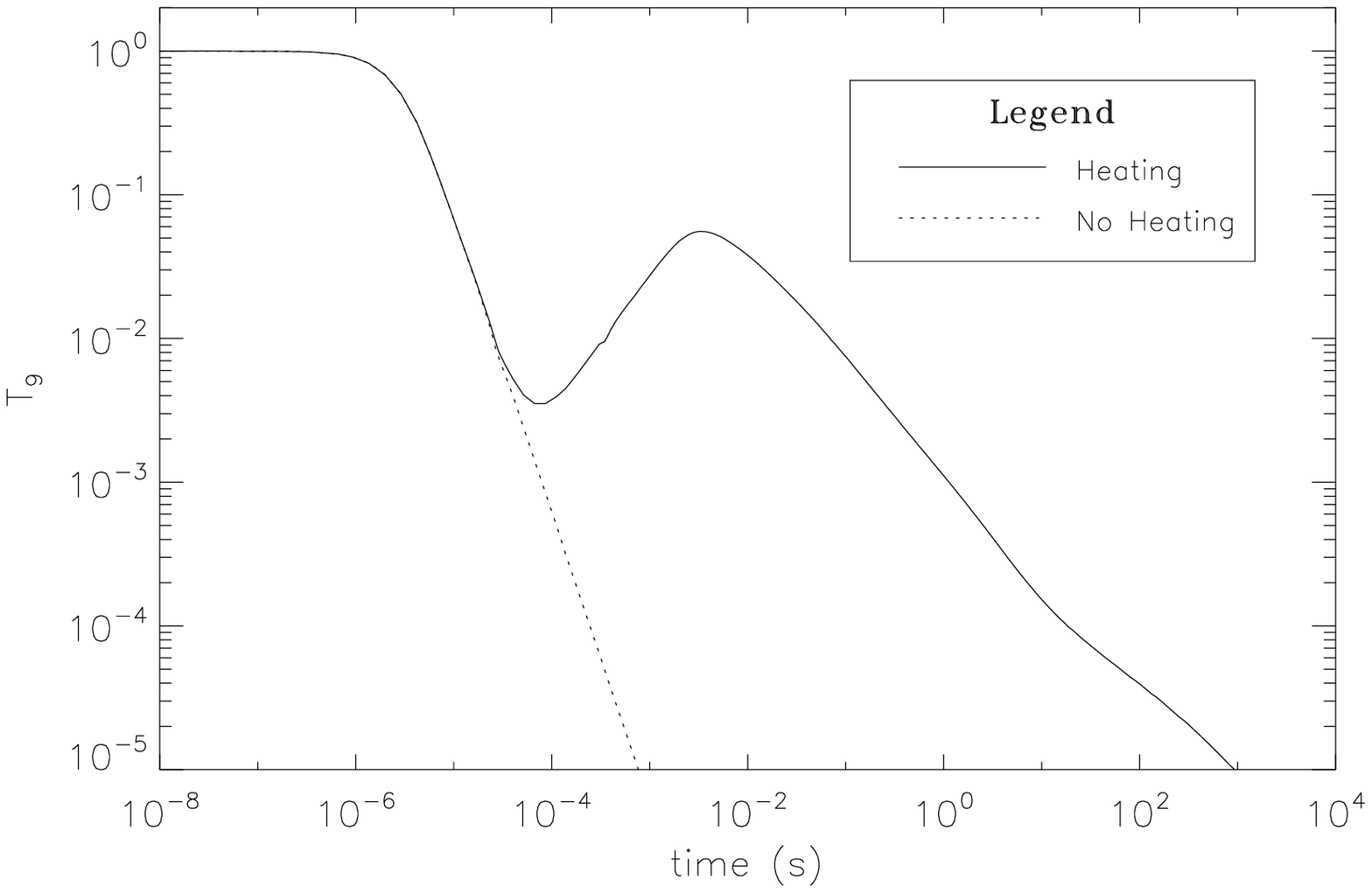}
}
\caption{
The temperature as a function of time for the $\alpha = 1$ expansion
in the cases in which heating from nuclear
reactions is included (Heating) or not (No Heating).
}
\label{fig:t9_alpha=1}
\end{figure}

\begin{figure}[ht!]
\centerline{
  \includegraphics[width=0.5\textwidth,angle=0]{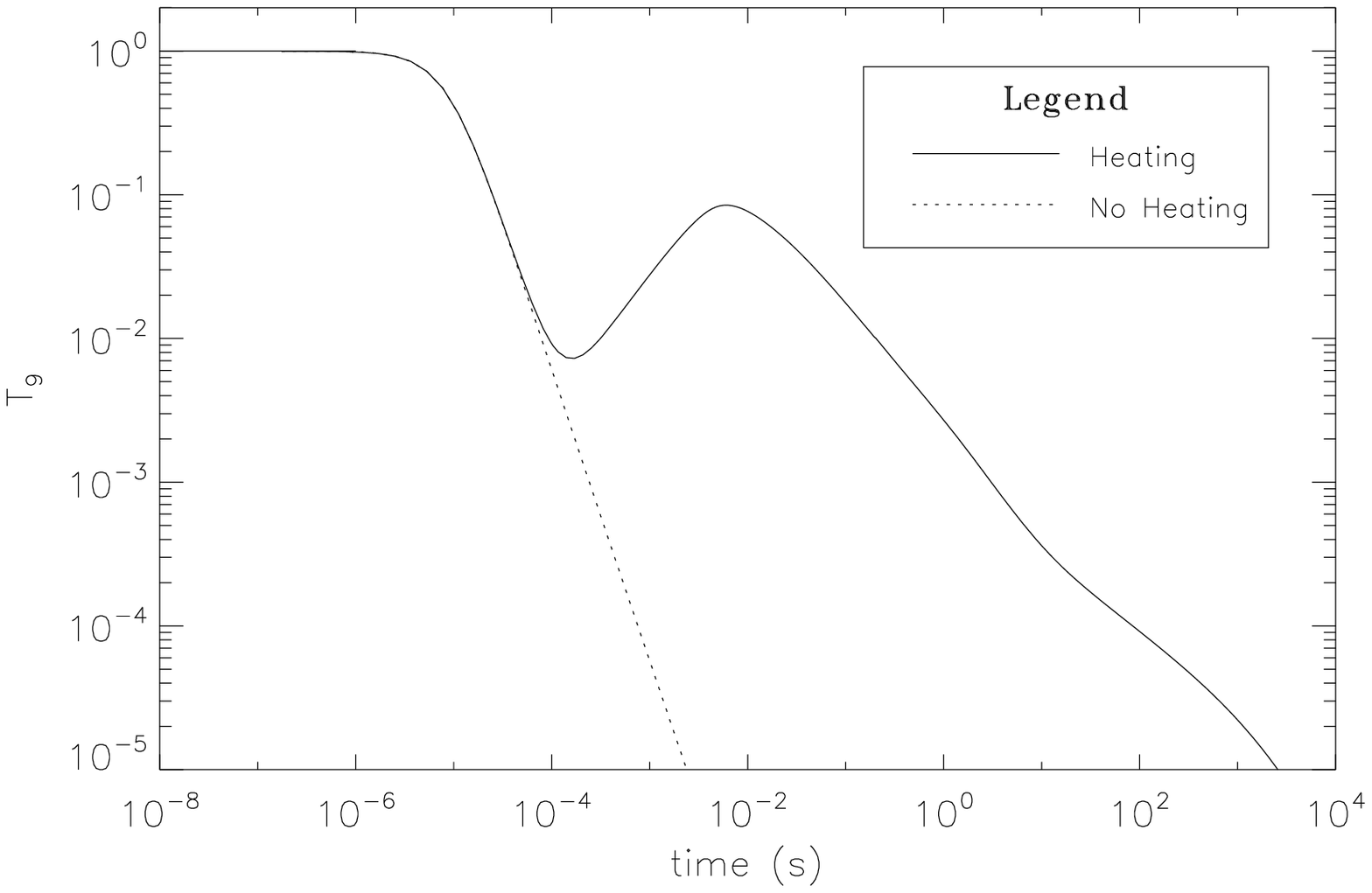}
}
\caption{
The temperature as a function of time for the $\alpha = 0.1$ expansion
in the cases in which heating from nuclear
reactions is included (Heating) or not (No Heating).
}
\label{fig:t9_alpha=0.1}
\end{figure}
\vskip 0.2cm
\noindent\underline{sensitivity to heating}:As is apparent, calculations that neglect heating from nuclear reactions reach higher mass nuclei.  The reason is evident from Figures \ref{fig:t9_alpha=1} and \ref{fig:t9_alpha=0.1}, which show the
time-dependence of the temperature for the various calculations.  When
heating is included, the temperature initially falls but then begins to
rise beginning around $10^{-4}$ seconds as some of the nuclei begin to
beta decay. The matter heats up, and the temperature
peaks at about $10^{-2}$ seconds.
At this time, the nuclei have reached the $N=82$ closed shell, and the
extra binding forces the r-process path closer to beta stability to the
so-called ``waiting-point'' nuclei where beta
rates are slower.  The heating thus becomes less effective.  In fact the
slow beta decays at the waiting-point nuclei
hold up the flow so much that the rapid
expansion allows the neutron-capture reactions
to freeze out before consuming the neutrons,
and the final abundance distribution is centered on the $N=82$ closed
shell.  This results in an r-process abundance
distribution peaked near $A=130$.

\vskip 0.2cm

\noindent If heating is not included, however, the temperature quickly falls to such
low temperatures that neutron disintegration reactions become ineffective.
The r-process flow pushes beyond the r-process waiting-point nuclei to
isotopes with larger beta decay rates.  The r-process flow can then reach
higher nuclear mass.  The rapid expansion still causes the matter to freezeout
before all neutrons are absorbed into nuclei, nevertheless, there is
considerable fission cycling.  Remarkably, the final r-process abundance
pattern does not look like the solar pattern.
Because the r-process flow pushes neutron rich
of the waiting-point nuclei, the extra binding of the closed neutron shells
does not impose itself on the final abundance distribution. Interestingly, the slower expansion with $\alpha=0.1$ does not look significantly different from the $\alpha=1$ case.  The flow does reach slightly higher mass in the case with full heating because some nuclei are able to leak past the $N=82$ closed shell before neutron-capture
reactions freeze out.  Nevertheless, the abundance distribution is still
dominated by the $A \approx 130$ peak.  It is clear that for $\alpha$ not
too different from unity, the degree of energy deposition has a larger effect than the expansion rate.

\vskip 0.2cm

\begin{figure}[ht!]
\centerline{\includegraphics[width=0.5\textwidth,angle=0]{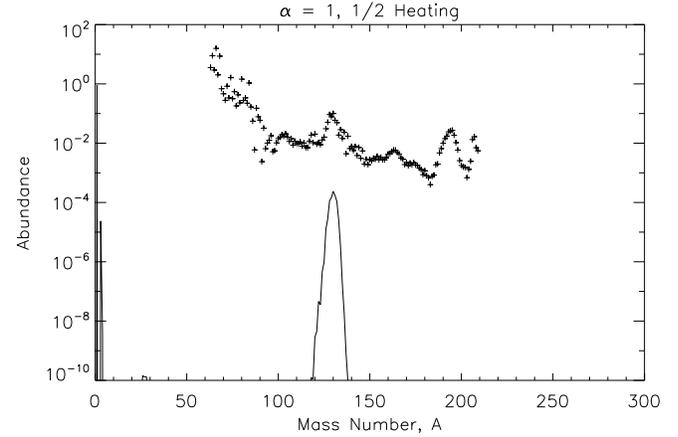}}
\caption{
 The final abundance distribution as a function of mass number for
 the \protect{$\alpha = 1$}
 calculation in the case in which 50\% of the energy released by
 nuclear reactions was assumed to be lost from the matter.}
\label{fig:ya_alpha=1_half_heat}
\end{figure}

\noindent We investigate a more realistic treatment of heating by
running a calculation assuming only 50\%
of the energy released by nuclear reactions is deposited locally (to
account for energy lost by neutrinos).  The final abundance pattern shown in Figure \ref{fig:ya_alpha=1_half_heat} is little different from that in the full heating calculation shown in Figure \ref{fig:ya_alpha=1_heat}. Although the expansion with half energy deposition reheats to a lower temperature than the calculation with the full heating, the resulting temperature is enough to allow
disintegration reactions to prevent the flow from bypassing the waiting
point nuclei.

\begin{figure}[ht!]
\centerline{\includegraphics[width=0.5\textwidth,angle=0]{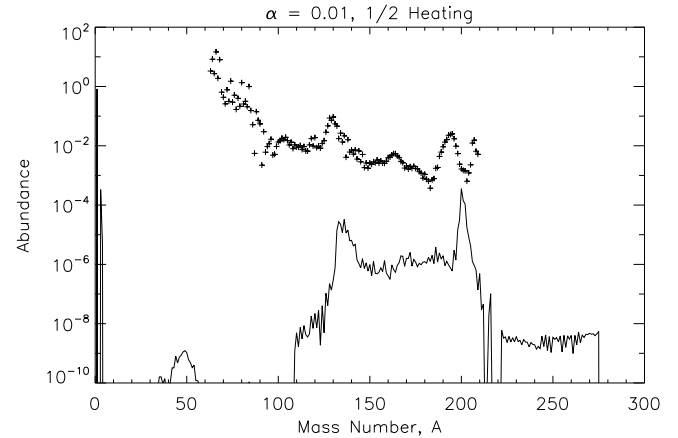}}
\caption{
 The final abundance distribution as a function of mass number for
 the \protect{$\alpha = 0.01$}
 calculation and 50\% local energy deposition from
 nuclear reactions.}
\label{fig:ya_alpha=0.01_half_heat}
\end{figure}

\begin{figure}[ht!]
\centerline{\includegraphics[width=0.4\textwidth,height=9.5cm,angle=270]{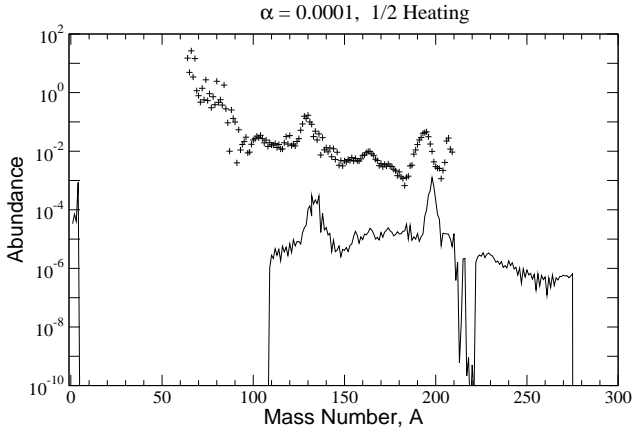}}
\caption{
 The final abundance distribution as a function of mass number for
 the \protect{$\alpha = 0.0001$}
 calculation and 50\% local energy deposition from
 nuclear reactions.}
\label{fig:ya_alpha=1e-4_half_heat}
\end{figure}

\noindent \underline{sensitivity to $\tau$}: The r-process yield is quite sensitive to $\tau$. To test this sensitivity, we study cases with $\alpha=0.01$ ($\tau\sim 0.1$ms) and $\alpha=0.0001$ ($\tau\sim 1$ms) in which we allow 50\% of the energy released by nuclear reactions to escape from the matter. The initial conditions in the latter case are similar to the choice made in Goriely et al. (2004). The final abundance distribution
is shown in Figures \ref{fig:ya_alpha=0.01_half_heat} and \ref{fig:ya_alpha=1e-4_half_heat}.  The relatively
slow expansion in these cases allows the nuclear flow to pass the $N=82$
waiting point nuclei before neutron-capture reactions freeze out.  These
expansions are the most promising in terms of reproducing the heavy solar
r-process distribution, as concluded by Goriely et al. (2004). 
Figures \ref{fig:ya_alpha=1_half_heat}, \ref{fig:ya_alpha=0.01_half_heat} and \ref{fig:ya_alpha=1e-4_half_heat} clearly portray the strong sensitivity of the yield to the expansion timescale $\tau$, which scales as $1/\sqrt{\alpha}$ (eqn.(3) of sec 4) in our simplified dynamical model of pressure-driven expansion. Smaller $\alpha$'s imply a slower expansion. In the absence of a hydrodynamical model for the explosion, we have explored a plausible range of timescales $0.01<\tau({\rm ms})<0.1$ by demanding that the mass ejection be completed before the entire star converts to quark matter. The principal effect of a large expansion timescale is that it counteracts the effect of heating from $\beta$-decays which halts the flow at the 2nd r-process peak and inhibits further neutron capture. A slower expansion can allow the flow to break past the 2nd peak before all neutrons freeze out. It is clear that the final abundance distribution is quite sensitive to the exact nature of the dynamical trajectory of the expansion and that detailed models that include a full treatment of the reheating from nuclear reactions will be needed to further elucidate the r-process in decompressing matter.

\noindent \underline{sensitivity to $Y_e$}: Since we lack a dynamical picture of the quark-nova at present, the amount and composition of ejected material is subject to uncertainty. Prior estimates of mass ejection by a quark-hadron phase transition suggest values ranging from 0.001 solar mass to 0.1 solar mass~(\cite{Fry}). The $Y_e$ will be
different in each case depending on the ejected mass. Therefore, we study the dependence of the yield pattern on a mass-averaged electron fraction $\langle Ye\rangle$, given by

\begin{equation}
\langle Ye\rangle = \frac{\int_{R_c}^{R_0}Ye(\rho)\rho dV}{\int_{R_c}^{R_0}\rho dV}
\end{equation}

where $R_c$ is chosen to give a particular total ejected mass (for the density profile as described below), and $R_0=10$km. 

\vskip 0.1cm

\noindent To determine $\langle Ye\rangle$, we choose the density profile of the neutron star as given in~\cite{LRP}. This profile coresponds to the BPS equation of state~(\cite{BPS}) at low density, matched to the BBP equation of state~(\cite{Baym}) at densities upto nuclear saturation density. We focus only on 0.01, 0.001 solar mass ejecta since there is considerable uncertainty for larger ejected mass coming from the high density equation of state. We also study $10^{-5}$ solar mass ejecta as an aside although such a low value is not feasible for the energetics of a quark-nova. Using a parameterized formula for the density dependence of Ye extracted from Table 1 in~\cite{Meyer89}, we find the mass-averaged $Y_e$'s to be: $\langle Ye\rangle$=0.03 (0.01 solar mass), $\langle Ye\rangle$=0.09 (0.001 solar mass), $\langle Ye\rangle$=0.12 ($10^{-5}$ solar mass) and perform r-process calculations for each of these cases. The following trends for the $\langle Ye\rangle$-dependence of the yield (with other parameters $\tau$ and amount of heating fixed) are apparent from Figure~\ref{fig:y_ealpha=1e-4_half_heat} which shows results for the smallest and largest $\langle Ye\rangle$:

\begin{figure}[ht!]
\centerline{\includegraphics[width=0.4\textwidth,height=9.5cm,angle=270]{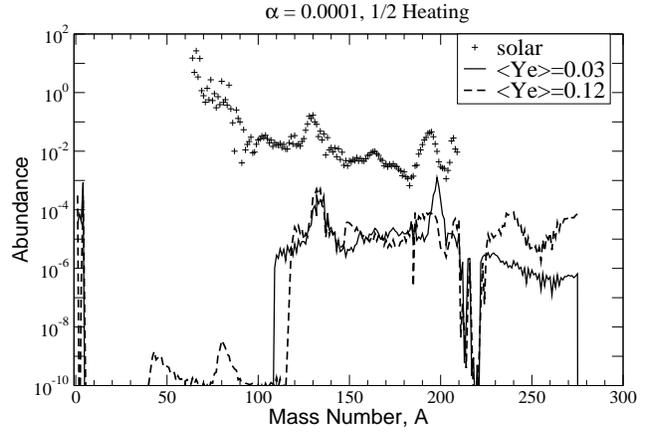}}
\caption{
 The final abundance distribution as a function of mass number for the
 \protect{$\alpha = 0.0001$}
 calculation, 50\% local energy deposition from
 nuclear reactions, and $\langle Y_e\rangle$'s as indicated.}
\label{fig:y_ealpha=1e-4_half_heat}
\end{figure}\vskip 0.1cm

\noindent 1) The yield of the actinide elements relative to lighter elements increases as  $\langle Ye\rangle$ increases\\
2) The position and shape of the second r-process peak is robust in all cases, while the third peak becomes less prominent and less stable with respect to the 2nd peak as $\langle Ye\rangle$ increases.\\
3) Smaller peaks appear around mass numbers $A\sim 44$ (mainly from heavy Ca, Ar and Ti  isotopes) and A$\sim 80$ (mainly from Se and Kr and their isotopes) as $\langle Ye\rangle$ increases. 
\vskip 0.1cm

\noindent These features show that in the quark-nova scenario, as in more conventional sites of the r-process, the 2nd and the 3rd r-process peak form under very different conditions of neutron exposure. The 2nd peak does not set very strict limits on the neutron exposure or the energetics of the quark-nova, but the 3rd peak requires significantly more mass ejection and smaller $\langle Ye\rangle$, which implies a more energetic explosion. For the case of slow expansion that we studied in these runs, the large neutron-to-seed ratio for small $\langle Ye\rangle$ depletes material beyond the 3rd peak due to strong fission cycling and depletes material below the 2nd peak due to rapid neutron capture. This depletion effect is weakened as we go to higher $\langle Ye\rangle$'s, hence the appearance of the smaller peaks, including the expected 1st r-process peak around Selenium 80. The dependence on $\langle Ye\rangle$ is therefore quite severe and deserves further study to constrain the energetics of, and element production in, the quark-nova. A coupled hydrodynamic and nucleosynthetic description of the quark nova needs to be done to better constrain the range of $Y_e$ and $\tau$.

\section{Quark-nova as an r-process site}

As observed in the results of the previous section, decompressing neutron matter can
be a site for heavy element production. Once the abundance pattern from such an
event has been generated, it is natural to turn to the broader
astrophysical questions of galactic enrichment by these events, and
consistency with observations detailed in the introduction. Though we
do not tackle this issue quantitatively in this work, we would like to
examine here some qualitatively important differences between quark-novae (assuming they power the decompression) and NSMs or type II SNe in relation to present observations of chemical abundances.
\subsection{QNe versus NSMs}
{\it \underline{Failure of NSE}}: 
\noindent The QN and NSM scenarios both utilize decompression of neutron matter
for the r-process. In case of NSMs, the ejected blobs of neutron matter
are located in the long spiral arms generated by the merger, while the
QN ejecta is assumed to flow outwards from the dense surface of a neutron star. 
Consequently, the density drops faster in case of NSMs than in the QN
scenario, and as a result, $\beta$-decays are more effective at heating up the material to nuclear statistical equilibirium (NSE) temperatures (\cite{Freibur}). 

\vskip 0.2cm

\noindent In the QN case, the slower expansion of the ejecta implies
that $\beta$-decays happen at higher densities (significant number of
$\beta$ decays occur when the $\beta$-decay timescale becomes
comparable to the expansion timescale) so that the temperature rises
at most to $T_9\sim 0.1$ (see Figures \ref{fig:t9_alpha=1} and \ref{fig:t9_alpha=0.1}) and NSE does not hold. 

\vskip 0.2cm

\noindent {\it \underline{r-process yield}}: 
\noindent The r-process yield from QNe are consistent with the total r-process yield in the Galaxy ($\approx 10^4M_{\odot}$, \cite{Waller}). Assuming that the phase transition to ($u,d,s$) matter happens due to spin-down of rotating neutron stars and accretion from the ISM, statistics on pulsar velocities imply that QNe occur at a rate of $f_{\rm QN}\approx 10^{-4}$-$10^{-6}$yr$^{-1}$ Galaxy$^{-1}$ (\cite{YHE})\footnote{The large range reflects the uncertainty in the high density equation of state, which is poorly known. For example, in Eriguchi et al. (2004), the bag constant B for quark matter is varied from $B=60$-$120$ MeV fm$^{-3}$. This affects the critical density at which the phase transition happens.}. Each QN event can release anywhere from $M_r=0.001$-$0.1M_{\odot}$ of baryonic matter for r-processing, therefore $M_{\rm tot}\sim f_{\rm QN}~M_r~t_{\rm age}$ is certainly in the range of our Galaxy's total r-abundance, given its age $t_{\rm age}\sim 10^{10}$ yrs.

\vskip 0.2cm

\noindent The coalescence rate for binary neutron stars is estimated to lie in the range $3.10^{-4}-10^{-6}$/y . Simulations of NSMs predict that, on average, about $10^{-2}M_{\odot}$ of baryonic matter is ejected per event. Once again, this is
consistent with our Galaxy's total r-abundance. The underproduction of elements at $A<130$ is a common feature between the two scenarios, but this depletion is
more pronounced in NSMs, since not all neutrons are consumed in the QN. 

\vskip 0.2cm

\noindent {\it\underline{Timescales}}:
 \noindent QNe can occur over widely different timescales. Once the central pressure inside a neutron star reaches a critical pressure, nuclear matter is in a metastable state where a droplet of quark matter is likely to be formed. This process is not immediate, however, since quantum nucleation must take into account the surface and curvature energy costs in making a quark matter bubble (\cite{Bombaci}). The formation of such a bubble with a critical radius determined by minimizing the Gibbs free energy can be described by a tunneling process, whose probability is exponentially sensitive to the central pressure. Furthermore, the nucleation time is then given by

\begin{equation}
\tau=(\nu_0p_0N_c)^{-1}
\end{equation}

\noindent where $\nu_0=10^{23}$ /sec is the number of collisions made by a subcritical (virtual) droplet of quark matter confined in a potential well, $p_0$ is the tunneling probability and $N_c=10^{48}$ is the number of droplet centers in the star's center (\cite{Iida}). Results from Bombaci et al. (2004) show that the nucleation time can vary by several orders of magnitude from 1 yr to Hubble times for small ($\sim 1\%$) changes in stellar mass (and hence pressures, for a given equation of state).

\vskip 0.2cm

\noindent Coalescence timescales for neutron star mergers in the majority of binary systems range from 100-1000 Myr, although several recent works argue for a significant population of short lived binaries with merger times less than 1 Myr~(\cite{Iva,Dewi,Kim}). This puts them at odds with the observed r-enhancement in some ultra-metal poor halo stars. This would not be a problem for QNe, which could happen in the very early stages of the universe's history since the phase transition can happen even within a year of forming a neutron star in a supernova event.

\vskip 0.2cm

\noindent{\it\underline{Chemical inhomogeneities}}:
As argued previously, NSMs are thought to be disfavored as dominant
astrophysical sites for the r-process since they would lead to a sudden and late
r-enrichment of the interstellar medium (ISM), leading to
chemical inhomogenities that are inconsistent with the observed
scatter in [r/Fe] of 0.2-0.3 dex (for possible caveats to this
argument, see Rosswog et al. (2000)). The underlying problem
is the large mass ejection in NSMs (hydrodynamic simulations indicate
that up to $10^{-3}$-$10^{-2}M_{\odot}$ of r-process matter might be
ejected in a merger event (\cite{janka99,ros04}, \cite{Rossw},2000)) as well as the low NSM rate (1-100 Myr$^{-1}$~Galaxy$^{-1}$,
with the upper limit being preferred). If we adopt the results of
Fryer \& Woosley (1998) on mass ejection from phase transitions, the QN can
eject as much material as an NSM, and their event frequency is also comparable. 
So, can the QN evade the problems faced by NSMs?

\vskip 0.2cm

\noindent If QNe are the primary site of the r-process, the constraints from metal-poor stars are difficult to satisfy. One would have to arrange for differing effective contributions to the r-process from QNe today than in the early history of the Galaxy. This might happen in the following way: Early in Galactic history, the distribution of gas (and therefore star formation) was more isotropic, so that the r-process material from QNe was absorbed into the next generation of stars. More recently, since most of the gas and star-forming regions came to reside in the Galactic disk, Type II SNe go off mainly in the disk. The
neutron stars thus created, on account of pulsar kicks, accquire large peculiar
velocities\footnote{It is a well known fact that pulsars in our Galaxy
have velocities much in excess of ordinary stars (\cite{Harrison})}, with transverse speeds ranging from $0$ to $\sim 1500$ km s$^{-1}$. This implies a mean three-dimensional speed of
$450\pm 90$ km s$^{-1}$ (\cite{LL}), hence some of these
run-away pulsars could cover kpc distances from their origins within a
Hubble time before they undergo a QN explosion as a result of an
increase of their core densities following spin down or accretion from
the ISM. Therefore, we expect that a percentage of the total r-nuclei
mass contributed by relatively recent QNe to reside in the diffuse gas around the halo of our Galaxy. Since the halo is not an active star-forming region, the r-process ejecta
is not absorbed into young stars, mitigating the problem of large chemical inhomogeneities
created by QNe. 

\vskip 0.2cm

\noindent 
It remains to perform a chemical evolution study of the enriched ISM from QNe events, and correlate the results with the observed scatter in elemental abundances. This is a further step in the modeling that will test the feasibility of the QN as an r-process site. At this stage, a detailed hydrodynamical description of the QN is lacking, and so a self-consistent analysis of chemical evolution cannot be performed; however, it is encouraging that the r-process yields calculated in this paper reveal the QN to be a candidate for an r-process site. Since the frequency of QNe in the Galaxy is similar to that of NSMs, we prefer to think of them as additional r-process sites to SNe, or even as substitutes for some failed/delayed SN events. 

\vskip 0.2cm

\subsection{QNe versus Type II SNe}
{\it\underline{Yields and timescales}}:
QNe can synthesize significantly larger amounts of
r-process material than the typical $10^{-6}-10^{-5}M_{\odot}$ thought
to be ejected in each core-collapse SN event (\cite{Cowan91};
\cite{Woosley}; \cite{Qian00}; \cite{Wanajo01}). The frequency
of QNe is less than that of SNe ($10^{-2}$~yr$^{-1}$ Galaxy$^{-1}$). Several
 of the uncertainties that plague the SN scenario also apply to the QN. The correlation of
r-process yields in SNe with progenitor mass is a complex problem,
while the size and density of the quark core in QNe is a crucial
factor controlling the ejected mass fraction. 

\vskip 0.2cm

\noindent {\it\underline{Role of neutrinos}}:
\noindent Neutrino propagation and absorption rates in hot and dense quark matter are not known as accurately as required, a problem that is exacerbated by the fact that even the
ground state of quark matter at high density is
uncertain\footnote{In the QN, we assume it to be three-flavors of free
massless quarks. In reality, a transition to a gapped phase is likely,
which can drastically alter the neutrino rates.}. The role of neutrinos in SNe and QNe is different, although the energy released in the form of neutrinos from core-collapse (in the case of SNe) and the phase transition (in QNe) is about the same.

\vskip 0.2cm

\noindent In SNe, neutrino-heating at the surface of the neutron star is
responsible for the neutrino-driven wind mechanism, and neutrinos
continue to be important for the r-process, especially in driving
fission of progenitor nuclei above $A\geq 190$ towards lighter nuclei
above and below $A\sim 130$.  Neutrinos can also induce
neutron emission accounting for the solar r-abundances at $A=183-187$
from the progenitor peak at $A=195$, and alter the neutron-to-seed
nuclei ratio (\cite{Haxton97}).
 
\vskip 0.2cm

\noindent In QNe, neutrinos deposit their energy {\it within} the
neutron star, and emerge as a neutrino burst only during the latter stages of r-processing. This justifies our neglect of neutrinos in the reaction network and their impact on the $Y_e$ of the ejected material during the initial stages of expansion. However, neutrino-induced effects as stated above will also be important in determining the final yield from a QN.

\vskip 0.2cm

\noindent{\it\underline{Compact remnants}}: An intriguing difference is that the compact remnant of type II SNe are neutron stars, whereas a QN will give birth to a quark star. If QNe do indeed occur in the universe, quark stars should abound, though not as much as neutron stars. A rough estimate of their number can be obtained by noting that our Galaxy likely contains about $10^8$ neutron stars, so that assuming QNe to be a primary site of the r-process, the constraint on the total r-process material in the Galaxy implies that 1 out of every 1000 neutron stars has undergone a QN. Since a QN leaves behind a quark star, this means that out of the approximately 1500 neutron stars observed so far, one or two may actually be quark stars. If QNe are only a secondary site, however, the above estimate for the number of quark stars may be highly optimistic, explaining why we have not detected any quark star as yet. This points to the need to distinguish clearly between neutron and quark stars, a distinction that can perhaps only be made by detailed observations of surface luminosities (\cite{Page}). While stellar mass and particularly radius measurements are at present not accurate enough to differentiate neutron and quark stars (\cite{Alf06}), recent studies of the thermal and non-thermal emission properties of the surface of bare quark stars point to super-Eddington luminosities of gamma-ray photons and $e^+e^-$ pair-winds as a unique and identifiable signature (\cite{Aksenov};\cite{Jaikumar}). However, the distinction may be undermined if quark stars have a crust (\cite{Jaikumar05}; \cite{Alford}). The signature of a QN with a fall-back crust was recently examined by Niebergal et al. (2006). The most direct test of the QN association to the r-process would come from $\gamma-$ decays of unstable nuclei produced in the r-process, as explained below.

\vskip 0.2cm

\noindent{\it\underline{$\gamma$-decay of r-nuclei}}:
\noindent We expect the QN ejecta to achieve $\gamma$-ray transparency sooner than SN ejecta since QN progenitors (i.e., neutron stars) lack extended atmospheres, so that r-process only nuclei with $\gamma$-decay lifetimes of the order of years (such as $^{137}$Cs, $^{144}$Ce, $^{155}$Eu and $^{194}$Os) can be used as tags for the QN. For example, given the r-process abundance of the above nuclei in the QN ($Y_e=0.03, \alpha=0.0001$, 1/2 heating), we can estimate the $\gamma$-flux as (\cite{Qian98})

\begin{eqnarray}
\frac{F_{\gamma}}{10^{-7}~\gamma~{\rm cm}^{-2}~{\rm s}^{-1}}&=&32\times I_{\gamma} \nonumber \\
&&\times\frac{\delta M}{10^{-7}M_{\odot}}\frac{100}{A}\frac{1{\rm yr}}{\tau}\left(\frac{10{\rm kpc}}{d}\right)^2    \label{gamm}
\end{eqnarray}

\noindent where $I_{\gamma}$ is the number of monochromatic photons per decay of the decaying nucleus with mass number $A$ and lifetime $\tau$, and $d$ is the distance to the QN. $\delta M$ is the expected mass of the $\gamma$-active r-process nucleus produced in a QN, which is estimated by assuming that a total of $10^{-2}M_{\odot}$ of r-process material is ejected per QN event. For the nuclei mentioned above, the value of $F_{\gamma}$ (in the units indicated by Eq.\ref{gamm}) is tabulated below for QNe, and compared to estimates from SNe (values taken from \cite{Qian98}):

\vskip 0.4cm

{\centering \begin{tabular}{|c|c|c|c|c|c|}
\hline
 $F_{\gamma}$ & $^{137}$Cs & $^{144}$Ce & $^{155}$Eu & $^{194}$Os & $^{126}$Sn
\\ \hline \hline
QN & $0.45$ & 2.0 & $0.92$ & $1.5$ & $0.30$\\ \hline
SN & $0.35$ & 2.9 & $0.50$ & $1.6$ & $5.0$\\ \hline
\end{tabular}\par}

\vskip 0.4cm \noindent In the above table, for cases where a nucleus displays
several gamma-ray transitions, we chose the one with the largest flux.
 For each of the short-lived nuclides (all except $^{126}$Sn), we chose a typical distance of $d\sim 10$ kpc for the r-process source.
For the case of $^{126}$Sn, which is a long-lived nucleus ($\tau=1.44\times 10^5$yr),
only fluxes from nearby sources would be detectable, therefore, we chose $d\sim 200$ pc. Even so, the flux from $^{126}$Sn decay is small, since the QN is not effective at producing elements under $A\sim 130$. We note that the line sensitivities implied by these fluxes are somewhat below the detection threshold capabilities of the
spectrograph (SPI) aboard the INTEGRAL satellite (\cite{Knodlseder}). Further, given the remaining lifetime of the INTEGRAL mission, and the rarity of QNe, it is advisable to look at $\gamma$-decays of long-lived radioactive nuclei such as $^{126}$Sn. 
However, unless a QN has occured in our vicinity ($d\leq 50$ pc) recently, gamma-ray fluxes from $^{126}$Sn decay would be too small to detect. We do not address
any possible connection with diffuse $\gamma$-emission at the galactic center since the frequency of QNe is quite small compared to the lifetime of $^{126}$Sn. The gamma decay of certain radioactive nuclides beyond $^{208}$Bi can also provide detectable signatures of an r-process site, provided the r-process event occurred nearby ($d\sim 0.2-1$ kpc) and recently ($t\leq$ 700 yrs). Expected fluxes from some promising nuclei identified in previous works (\cite{Clayton}, \cite{Qian99}) are calculated for a QN and the results are tabulated below, along with comparisons to the flux from a SN event.

\vskip 0.4cm

{\centering \begin{tabular}{|c|c|c|c|c|}
\hline
 $F_{\gamma}$ & $^{226}$Ra & $^{229}$Th & $^{227}$Ac & $^{228}$Ra 
\\ \hline \hline
QN & $1.7$ & 0.34 & $1.42$ & $3.24$\\ \hline
SN & $1.0$ & 0.21 & $1.2$ & $3.8$\\ \hline
\end{tabular}\par}

\section{Conclusion}

The r-process remains a complex nucleosynthetic process,
requiring extensive modeling of the astrophysics as well as nuclear
aspects. One of the underlying difficulties is that we are yet to
identify the astrophysical site in which ideal r-process conditions
are met. In this work we suggest that decompressing neutron star matter, triggered by an explosive phenomenon like a quark-nova may be a feasible
site. Decompressing neutron star matter provides a rich supply of exotic nuclei and large neutron flux, and generates an r-process environment that is similar, though not identical, to neutron star mergers. We performed network calculations adapted to the study of the decompression and the subsequent r-process, and found that final r-process yields and abundance patterns are determined primarily by dynamical timescales characteristic of a quark-nova (or the underlying motive force behind the decompression), and by the extent of heating from nuclear reactions. In the context of reproducing the observed pattern of solar abundances, the slowest expansions are the most promising. Rapid expansions, with or without nuclear heating, lead to a prominent 2nd peak but do not resemble the solar pattern. As a candidate for the r-process site, quark-novae can possibly evade some of the constraints from abundance observations of r-process enhanced metal-poor stars that seem to disfavor neutron star mergers as primary r-process candidates. On the basis of a preliminary exploration of the r-process yields, and qualitative features of the quark nova, we find it to be a promising scenario, to be developed and examined in more
detail. A numerical study of the dynamics of the quark-nova needs to be performed to relate the mode of burning of the nuclear-quark conversion front to the expansion velocity of the ejecta; here we have only explored a plausible regime of expansion
timescales. Finally, in our picture, quark-novae are intimately connected with the birth of
strange stars, so we suggest that decays of $\gamma$-active r-process nuclei be used as a tag for strange stars. This intriguing link between strange stars and r-process nucleosynthesis might be confirmed by the next generation of gamma-ray satellites.
\section*{Acknowledgments} 
The authors are grateful to Jim Truran, Amanda Karakas and Ken Nollett for comments on the text. P.J. acknowledges helpful discussions with Lee Sobotka and Mark Alford, and support for this work by the Department of Energy (DoE), Office of Nuclear Physics, contract no. W-31-109-ENG-38 and US DoE grant DE-FG02-93ER40756. R.O is supported by an operating grant from the Natural Research Council of Canada (NSERC) as well as the Alberta Ingenuity Fund (AIF). K.O. is supported by the National Science Foundation under grant PHY 02-16783 for the Physics Frontier Center Joint Institute for Nuclear Astrophysics (JINA). B. S. M. is supported by the Department of Energy through SciDAC and by grants from NASA's cosmochemistry program.

\end{document}